# The cosmological model of eternal inflation and the transition from chance to biological evolution in the history of life: The possibility of chance emergence of the replication and translation systems, and the protein superfolds

Eugene V. Koonin[*]

National Center for Biotechnology Information, National Library of Medicine, National Institutes of Health, Bethesda, MD 20894, USA

*For correspondence; email: koonin@ncbi.nlm.nih.gov

**Abstract**

Evolution of life on earth was governed, primarily, by natural selection, with major contribution of other evolutionary processes, such as neutral variation, exaptation, and gene duplication. However, for biological evolution to take off, a certain minimal degree of complexity is required such that a replicating genome encodes means for its own replication with sufficient rate and fidelity. In all existing life forms, this is achieved by dedicated proteins, polymerases (replicases), that are produced by the elaborate translation system. However, evolution of the coupled system of replication and translation does not appear possible without pre-existing efficient replication; hence a chicken-egg type paradox. The currently preferred solution is the concept of the RNA World which is conceived as a community of RNA molecules replicating without the help of proteins, with versatile catalytic activities including the replicase activity. However, despite considerable accumulated evidence of catalytic activities of RNA molecules, the RNA World concept encounters major hurdles and so far has offered no convincing scenarios for the origin of efficient replication and translation. I argue that the "many worlds in one" version of the cosmological model of eternal inflation implies that emergence of replication and translation by chance, as opposed to biological evolution, is a realistic possibility. Under this model, any life history that does not violate physical laws is realized an infinite number of time in the infinite universe although the frequencies of different histories are vastly different. Thus, the complex system of coupled translation and replication that is required for the onset of biological evolution would emerge an infinite number of times by pure chance although the probability of its appearance in any given region of the universe is vanishingly small. Furthermore, the emergence of the translation system would entail the origin of the major protein folds (the superfolds) in a big-bang-type event. After the chance emergence of the systems of replication and translation, and the major protein types, the transition from chance to biological evolution would occur. All the subsequent history of life in the infinite number of biospheres was determined by biological evolution, the principal law of biology that is an inevitable consequence of replication with a feedback loop. A major corollary of this scenario is that an RNA world, as it is currently conceived, might have never existed although catalytic activities of RNA were, probably, critical for the onset of biological evolution and its early stages.

> *The Library is unlimited and cyclical.* **If an eternal traveler were to cross it in any direction, after centuries he would see that the same volumes were repeated in the same disorder (which, thus repeated, would be an order: the Order).**
>
> **J. L. Borges,** *The Library of Babel,* **1951 [1]**
>
> **…the origin of protein synthesis is a notoriously difficult problem.**
> **F.H.C. Crick et al., 1976 [2]**

**Background**

*Comparative genomics and early stages of evolution: advances and limitations*

The recent advances in comparative genomics allow us to reach far back into early stages of cellular evolution. Identification of orthologous genes in diverse life forms and the use of appropriate reconstruction methods provides for mapping of gene origin to ancestral life forms. In particular, the gene repertoire of the Last Universal Common Ancestor (LUCA) has been, at least, partially reconstructed with a reasonable degree of confidence, suggesting that LUCA already contained at least a few hundred genes [3-5]. Comparative-genomic analyses and evolutionary reconstructions revealed remarkable differences between the evolutionary patterns of different functional systems. Whereas some of these systems, above all, the translation machinery and the core of the transcription system, are (nearly) universally conserved and hence confidently traced back to LUCA, other essential systems, in particular, those of DNA replication and membrane biogenesis, are highly diverged or even non-homologous in bacteria and archaea, the two primary kingdoms of prokaryotes [6-10]. This prompted the development of hypotheses that, despite its considerable complexity, LUCA might have been an entity qualitatively different from modern cells, possibly, with an RNA genome [7, 11, 12] and, perhaps, even not a membrane-bounded cell [8, 13]; with a new type of evidence, these evolutionary scenarios revive the older notion of LUCA as a progenote, a primitive life form[14]. Competing hypotheses envisage a LUCA that was similar to extant prokaryotic cells and emphasize replacement of even key functional systems, such as the DNA replication and the membrane during subsequent evolution [10, 15, 16].

Evolution prior to LUCA is a harder and more controversial area although, recently, scenarios for the emergence of the first cells have been developed, in particular, on the basis of the concept of ancestral life forms as communities of virus-like genetic elements [11, 13, 17]. Taken together, these findings and conceptual developments represent a considerable inroad into the early evolution of life. However, despite extensive experimental and theoretical efforts, there has been relatively little progress in explaining the origin of the genuine breakthroughs that must have occurred prior to the beginning of the bona fide biological evolution: efficient genome replication, the genetic code, and translation. Most often, the emergence of these central attributes of life is placed within the framework of the RNA World concept. However, the RNA World has its share of serious difficulties and so far has not been able to provide satisfactory accounts of the onset of biological evolution leading to the origin of replication and translation. Here I offer a radical alternative rooted in modern cosmological models.

**Hypothesis: on the possibility of a chance origin a complex system amenable to biological evolution in light of modern cosmological models**

*The central problem: the emergence of biological evolution, the Eigen limit, the limitations of the RNA world and the inherent paradoxes of the origin of replication and translation systems*

The origins of replication and translation (hereinafter ORT) appear to be qualitatively different from other problems in evolutionary biology, even those that involve deep mysteries, such as the origin of eukaryotes. As soon as sufficiently fast and sufficiently accurate genome replication emerges, **biological evolution** takes off. I use this unassuming term because, in addition to Darwinian natural selection[18], other evolutionary processes such as fixation of neutral mutations that might provide material for subsequent adaptation [19], exaptation of "spandrels" (recruitment of features that originally emerge as by-products of the evolutionary process but are subsequently utilized for new functions) [20], and duplication of genome regions followed by mutational and functional diversification [21]. All these processes that together comprise biological evolution become possible once sufficiently accurate and sufficiently fast replication of the genetic material is established. The required accuracy is defined by the so-called Eigen limit which follows from the theory of self-replicating systems developed, primarily, by Eigen and coworkers in the 1970ies [22]. Simply put, if the product of the error (mutation) rate and the information capacity (genome size) is safely below one (i.e., less then one error per genome is expected per replication cycle), most of the progeny are exact copies of the parent, and reproduction of the system is sustainable.

If this value is significantly greater than one, most of the progeny differ from the parent, and the system is headed for an error catastrophe (a term and idea traceable to the early concept of Orgel developed in the context of a possible contribution of translation errors to aging [23]). The Eigen limit would have been of particular importance to early, primitive organisms. Indeed, the very origin of the first organisms presents, at least, an appearance of a paradox because a certain minimum level of complexity is required to make self-replication possible at all, and high-fidelity replication requires additional functionalities that need even more information to be encoded. At the same time, the existing level of replication fidelity limits the amount of information that can be encoded in the genome[22, 24, 25]. What turns this seemingly vicious circle into the (seemingly) unending spiral of increasing complexity (the Darwin-Eigen cycle, according to Penny [26]) is natural selection conditioned on genetic drift. Even small gains in replication fidelity are advantageous to the system, if only due to the decrease of the reproduction cost as a result of the increasing yield of viable copies of the genome. In itself, a larger genome is more of a liability than an advantage due to higher replication costs. However, moderate genome increase, e.g., by duplication of parts of the genome or by recombination, can occur by genetic drift in small populations [27]. Replicators with fidelity exceeding the Eigen limit can take advantage of such randomly fixed and, initially, useless genetic material by evolving new functions, without falling off the "Eigen Cliff". Among such newly evolved, fitness-increasing functions will be those that increase replication fidelity which, in turn, allows further increase in the amount of encoded information. And so the Darwin-Eigen cycle perpetuates itself in a spiral progression, leading to a steady increase in genome complexity.

The crucial question on the origin of life is how did the Darwin-Eigen cycle start, i.e., how was the minimal complexity attained that is required to achieve the minimally acceptable replication fidelity. In even the simplest modern systems, such as RNA viruses with the replication fidelity of only ~$10^{-3}$, replication is catalyzed by a complex protein replicase; even if one disregards accessory subunits that are present in most polymerases, the main catalytic subunit is a protein that consists of at least 300 amino acids [28]. The replicase, of course, is produced by translation of the respective mRNA which is mediated by a tremendously complex molecular machinery. Hence the ***first paradox of ORT***: in order to attain the minimal complexity required for a biological system to get on the Darwin-Eigen spiral, a system of a far greater complexity appears to be required. How such a system could evolve, is a puzzle that defeats conventional evolutionary thinking, all of which is about biological systems moving along the spiral; the solution is bound to be unusual.

The commonly considered way out of this "Catch 22" is the proposal that replication evolved before translation, in an "RNA world" in which both the genomes and the catalysts were RNA molecules [29-31]. The RNA world idea was incepted as the result of the discovery of self-splicing introns and the catalytic RNA moiety of RNAse P [32, 33], the first ribozymes. In the elapsed 20 years, the RNA World has gained a lot of momentum[34]. In particular, diverse catalytic activities of ribozymes have been observed in natural RNA molecules or engineered experimentally [35-38]. However, all the advances of ribozyme studies notwithstanding, the prospects for a *bona fide* ribozyme replicase remain dim as the ribozymes designed for that purposes are capable, at best, of the addition of ~10 nucleotides to a oligonucleotide primer, at a very slow rate and with fidelity at least an order magnitude below that required for the replication of relatively long RNA molecules that are required to produce complex ribozymes, such as the replicase itself [39, 40]. To quote leaders in the field: "These experiments demonstrate that RNA replicase behavior is likely within the catalytic repertoire of RNA, although many obstacles remain to be overcome in order to demonstrate that RNA can catalyze its own replication in a manner that could have sustained a genetic system on the early Earth."[39]

The origin of the translation system might be an even harder problem that inevitably brings up *the second paradox of ORT*: high translation fidelity hardly can be achieved without a complex, highly evolved set of RNAs and proteins but an elaborate protein machinery hardly could evolve without an accurate translation system. The common solution, again, is offered by the RNA World concept in the form of a highly evolved RNA-only translation system. The role of RNA catalysis in modern and, potentially, to a greater extent, in ancient translation enjoys considerable experimental support. Indeed, the experimentally demonstrated activities of ribozymes include, among others, those that are involved in the main chemical steps of translation such as amino acid activation, RNA aminoacylation, and peptidyl transfer [41-44]. Moreover, it has been unequivocally shown that the large subunit rRNA itself is a ribozyme that catalyzes the peptidyl transferase reaction [45, 46]. Nevertheless, the problems with RNA-only translation remain serious. So far, no one has been able achieve translation without proteins, it remains unclear whether translocation, the principal mechanistic step of translation is mediated by RNA, and the rate and fidelity of translation after the removal of translation factors are quite low [47].

Perhaps, most dramatically, a scenario of Darwinian evolution of the translation system from the RNA world is extremely hard to conceive. Herein comes the ***third paradox of ORT***: until the evolving translation produces functional proteins, there is no obvious selective advantage to the evolution of any parts of this elaborate (even in its most primitive form) molecular machine. Conceptually, this paradox is, of course, closely related to the general problem of the evolution of complex systems that was first recognized by Darwin in his famous discussion of the evolution of the eye [18]. The solution sketched by Darwin centered around the evolutionary refinement of a primitive version of the function of the complex organ (photosensitivity in the case of the eye); subsequently, the importance of the exaptation route for the evolution of complex systems has been realized [20]. However, the problem of the origin of the translation system is conspicuously resistant to the application of either line of reasoning. Low-fidelity, slow translation, perhaps, even in a protein-free system, is conceivable as an intermediate stage of evolution but that does not solve the problem because, even for that form of translation, the core mechanism should have been in place already. Speculative scenarios have been developed on the basis of the idea that even short peptides could provide selective advantage to an evolving system in the RNA world by stabilizing RNA molecules, affecting their conformations or enhancing their catalytic activities ([48] and see the accompanying paper). This is compatible with observed effects of peptides on ribozyme activity [49] but it has to be admitted that none of these scenarios is either complete or compelling or supported by any specific evidence.

The origin of genetic coding is the ***fourth paradox of ORT***, and this is, perhaps, the most formidable one. In the modern translation system, the amino acid-codon correspondence is secured by the combination of base-pairing between the mRNA codons and the cognate tRNA anticodons, and specific charging of the cognate tRNAs with the respective amino acids which is catalyzed by specific aaRS. Evolution of such a mechanism is a mystery because it is unclear how would the specificity of tRNA charging evolve in the absence of aaRS. Despite considerable experimentation, there is no convincing evidence of specific affinities between amino acids and the cognate codons or anticodons [50, 51]. Thus, a completely unknown mechanism of RNA-amino acid recognition must have operated in the evolving primitive translation system.

The message of this section is not that ORT is a problem of "irreducible complexity" and that the path from monomers (nucleotides and amino acids) to the systems of replication and translation is not passable by means of biological evolution. It is fully conceivable that a compelling evolutionary scenario is eventually developed and, perhaps, validated experimentally. However, it is equally clear that the ORT problem is not just an extremely hard one but, arguably, the hardest problem in all of evolutionary biology. For all other problems, at least, the basal mechanism, genome replication, is available but, in the case of ORT, the emergence of this mechanism itself is the explanandum. Thus, it is of interest to consider radically different scenarios for ORT. I believe that modern cosmological models imply that such radical alternatives to biological evolution as the path to the replication and translation systems might be viable; this is discussed in the next sections.

*Evolution of the cosmos: eternal inflation, "many worlds in one" and the antrhopic*

The "many worlds in one" version of the cosmological model of eternal inflation makes a startling prediction: that all macroscopic, "coarse-grain" histories of events that are not forbidden by conservation laws have realized (or will realize) somewhere in the infinite universe, and not just once but an infinite number of times [52, 53]. According to Garriga and Vilenkin, "there are infinitely many *O*-regions where Al Gore is President and – yes – Elvis is still alive."[52] (*O*-regions, in the notation of Garriga and Vilenkin, are Observable regions of the universe like the one in which we live). Of course, under this model, the number of *O*-regions where The King is alive is vanishingly small compared to those where he was never born, and in the vast majority of these, there are no life, no humans, and even no stars and planets. The major conclusions, nevertheless, is that each history that is permitted in principle will be repeated an infinite number of times. To say that this worldview is counterintuitive seems to be a serious understatement.

However, to my knowledge, none of the publications that develop this model appeared on or around April 1, and it is seriously considered by the leading cosmologists [54]. Indeed, the "many worlds on one" concept is a direct consequence of eternal inflation, the dominant model of the evolution of the universe in modern cosmology [55-57]. The key insight of Garriga and Vilenkin is that the number of semi-classical, coarse-grain histories that can realize in an *O*-region is finite, even if astronomically vast (on the order of $10^{150}$)[53].

Inflation denotes the period of the exponentially fast initial expansion of a universe [58]. In the most plausible, self-consistent inflationary models, inflation is eternal, with an infinite number of island universes emerging as the result of decay of small regions of the primordial "sea" of false vacuum. For observers located within each island universe, their universe appears to be self-contained and infinite, and containing an infinite number of *O*-regions. From the point of view of such observers (like us), their (island) universe is expanding from a singularity (Big Bang) which corresponds to the end of inflation in the given part of the universe. The initial conditions at each big bang are determined by random quantum processes during inflation. Inflation is in excellent agreement with several crucial, recent results of observational cosmology - above all, the flatness of space in our *O*-regions, the overall uniformity of the cosmic microwave background radiation, and its local non-homogeneities[54]. Thus, although the model of eternal inflation cannot be considered proved, this is the strongly preferred current scenario of the cosmic evolution.

Garriga and Vilenkin realized that the content of each *O*-region can assume only a finite number of states and, accordingly, any *O*-region has a finite, even if vast, number of possible macroscopic, coarse-grain histories [52]. Garriga and Vilenkin present several lines of argument supporting this conclusion, which I will not retell here; in any case, the finiteness of the number of coarse-grain histories appears to be a straightforward consequence of the quantum uncertainty[53]. Combined, the model of eternal inflation and the notion that the number of coarse-grain histories is finite inevitably lead to the conclusion that each history permitted by conservation laws is repeated an infinite number of times (the quantum randomness of the initial conditions at the big bang is a pre-requisite of this conclusion; there seems to be no reason to question such randomness).

The many worlds in one model is tightly linked to the anthropic principle (or, more precisely and relevantly, ***anthropic selection***), which is a highly controversial, and yet, increasingly popular concept among cosmologists. Anthropic selection means that the only "reason" the island universe where we live has its specific properties (by this, cosmologists mean, primarily, the "fine-tuning" of the fundamental constants of nature, such as the masses of proton, neutron, and electron, the gravity constant etc) is that, should any of these properties be significantly different, we would not be here to peer into the universe[59, 60]. In the original formulation of Brandon Carter, "what we can expect to observe must be restricted by the conditions necessary for our presence as observers"[61]. The existence of a specific link between the fundamentals of the evolution of the universe and the origin of complex life forms is dubious at best, so in practice, anthropic selection pertains, more generically, to conditions conducive to the origin and long-term survival of life (this is the "weak" anthropic principle; the strong version has mystical reverberations and is better left alone). Within the "many worlds in one" model, anthropic selection has a clear, intuitive interpretation: the parameters of our *O*-region are selected among the infinite number of parameter sets existing in the universe by virtue of being conducive to the emergence of life. Under the "many worlds in one" model, there is no question whether or not other regions like that exist: they positively do, and moreover, their number is infinite. However, a most pertinent, indeed, a burning question is: how common or how rare are such *O*-regions in the universe?

The "many worlds in one" model changes the very notions of "possible" and "likely" with regard to various events and historical scenarios: indeed, any history, however complex but not violating the laws of physics, is not just possible but is, inevitably, realized in an infinite number of *O*-regions. However, the prior probabilities of the vast majority of histories to occur in a given *O*-region are vanishingly small. It is important to emphasize that the "many worlds in one" model is by no means a metaphysical conjecture but rather a straightforward implication of the eternal inflation cosmology that should be taken as literal physical reality [62]; of course, its validity is fully contingent on the validity of eternal inflation itself (see above). This distinctive worldview is bound to have profound consequences for the history of any phenomenon in the universe, and life on earth cannot be an exception[1]. These consequences are examined in the next section.

*The transition from chance to biological evolution in the history of life, the possibility of the origin of the core translation-replication system by chance, the no-RNA-World scenario, and the main features of the breakthrough system*

The "many worlds in one" version of the eternal inflation model implies a fundamentally different perspective on the origin of life. Under this worldview, a system of any complexity whose existence is permitted by the laws of physics can and does emerge in the infinite universe, and in an infinite number of copies. Importantly, however, the probabilities of different configurations of matter can be vastly different, and the specific question, with regard to life on our planet, is: what kind of scenario is most likely to apply? I examine this issue within the framework of the **chance to biological evolution transition** (Fig. 1). The existence and importance of this transition in the history of life seems to be beyond doubt, even if it is not often discussed. Indeed, biological evolution cannot possibly take off before there are polynucleotides (most likely, RNA molecules) and means for their sustainable replication. Thus, the synthesis of monomers (at least, nucleotides) and oligomers (short RNA molecules) could not have evolved biologically and must have been established by chance. At the other end of the spectrum, there is no doubt that the first fully fledged cells evolved via the biological evolution process. Somewhere in between lies the transition point, the threshold of biological evolution. Most often, since the advent of the RNA World concept, this threshold is (implicitly) placed in the RNA World and is associated with the emergence of replicating RNA molecules; the origin of translation is, then, assigned to a later stage of evolution and is thought to be brought about by an unspecified or, at best, invented *ad hoc* selective process (see discussion above and in the accompanying paper). Given the formidable difficulties encountered by all attempts to explain the ORT by means of (pre)biological evolution and prompted by the "many worlds in one" version of the eternal inflation model of cosmology, I suggest that the threshold might lie much higher on the axis of organizational complexity (Fig. 1). Specifically, I submit that the possibility that the breakthrough stage for the onset of biological evolution was the core of the coupled system of translation-replication, which emerged purely by chance, should not be dismissed, however counterintuitive. The "many worlds in one" model not only permits but guarantees that, somewhere in the infinite universe, such a system would come to be. The pertinent question is whether or not this is the likely breakthrough stage for the actual history of life on earth. I suggest that such a possibility should be taken seriously, given the problems encountered by all attempts to trace biological evolution to simpler systems. A central corollary to this hypothesis is that the RNA World, as it is currently pictured, i.e., a vast community of replicating RNA molecules possessing a variety of catalytic activities including that of an RdRp capable of replicating other RNA molecules in trans, but no translation system and no genetically encoded proteins, might have never existed. Of course, as discussed below, this does not at all rule out the special importance of ribozymes in early evolution, in particular, in the evolution of translation.

---

[1] This model and, indeed, its implications for the evolution of genomes seem to have been presaged with uncanny accuracy, some 50 years before its formulation, in Borges's masterpiece, The Library of Babel 1.
    Borges JL: **Collected Fictions** New York: Penguin; 1999.

The modern translation apparatus shows clear signs of evolution by duplication-diversification in the essential, ubiquitous components, allowing one to glean some features of the putative breakthrough system. It appears that the breakthrough system was an RNA-based machine, to a much greater extent than the modern translation system, vindicating a "weak" RNA world notion. Specifically, the aminoacyl-tRNA synthetases (aaRS) comprise two unrelated classes each of which evolved via a series of duplications[63, 64]. Moreover, both families of paralogous aaRS are relatively late elaborations within large classes of nucleotidases[65-67], strongly suggesting that the breakthrough system activated amino acids via an RNA-only mechanism. The same pertains to the translation factors that are relatively late products of evolution within the GTPase class of the P-loop NTPases; thus, the breakthrough system would not employ protein translation factors[68]. The phenomenon of so-called mimicking of tRNA structures by some of the translation factors [69-71] further supports the notion that the ancestral translation system was RNA-centered. Perhaps, most importantly, the rRNA itself is a ribozyme that catalyzes transpeptidation, the central chemical step of translation (see above). The most remarkable and enigmatic feature of the modern translation machinery is the common structure and the presence of some conserved sequence elements (e.g., the pseudouridine loop) in the tRNAs of all specificities which suggests that all the tRNAs are ancient paralogs[72]. Thus, the breakthrough system, conceivably, would utilize adaptors that were simpler than tRNAs, with the latter taking over already at the biological evolution stage.

The above considerations based on the comparative analysis of translation system components imply that the breakthrough system was a primitive translation machine that consisted solely of RNA and translated exogenous RNAs such that at least one functional protein, a generic RdRp, was generated by chance. The RNA viral RdRps, conceivably, the first replicative enzymes to emerge, possess a relatively simple domain structure related to that of a class of non-enzymatic RNA-binding proteins, the RRM[66]. Whereas, under the RNA world scenario, it would be inferred that the ultimate ancestor of this class of proteins was a non-enzymatic RNA-binding protein facilitating the action of a ribozyme replicase, the present scheme implies the primacy of the replicase, probably, one with a relatively low and non-specific activity. Once, in such a system, a functional RdRP is produced, making replication possible, biological evolution would kick off, including gradual "invention" of new proteins.

Evolution of the genetic code is another major aspect of the origin of biological organization. The code has been reported to be optimized with respect to mutational and, probably, also translational robustness, such that the universal code of modern life forms is more robust than $10^6$-$10^9$ random codes[73-75]. This robustness is manifest in the well-known non-randomness of the code structure such as the (partial) redundancy of the third base or the fact that all codons with a U in the second position encode hydrophobic amino acids[76]. This is normally viewed as a result of evolutionary optimization of the code[74]. However, the "many worlds in one" model is conducive to an alternative view under which, among the huge number of codes that emerged by chance in different regions of the universe, only codes with a certain, minimal robustness would allow the appearance of a functional replicase. At a new level and from a different perspective, this revives the old notion of the origin code by "frozen accident" discussed by Carl Woese and Francis Crick in the classical studies of the 1960ies[77, 78]. However, an important distinction of the view developed here is that the accident is not, actually, accidental, in the sense that a less robust code would not provide for the onset of biological evolution. This is, clearly, a case of anthropic selection (see above) which, in this case, would precede the onset of biological selection. This scenario for the emergence of the code does not preclude limited adjustments, codon reassignments, and capture of some codon series by new amino acids, phenomena that have occurred, to a small extent, during evolution of organelles and some parasitic bacteria[79], and have been suggested to play a more prominent role during early evolution[80].

To summarize, the scenario of the chance emergence of a coupled translation-replication system eliminates the first three paradoxes of ORT by postulating that replication and translation, in their most basic forms, actually, have not evolved. There is a connection to the RNA world concept in that an RNA-only translation system is postulated; however, the central tenet of the RNA world, the replication of a complex mixture by ribozyme replicases, is not a part of this scenario. Notably, however, the fourth paradox of ORT, that of the origin of the correspondence between amino acids and codons, is not solved. Even under this scenario, the breakthrough system would have an unknown, RNA-based mechanism for establishing this crucial correspondence, and from this primordial mechanism, the modern aaRS-based one must have evolved.

The general setting for the emergence of the breakthrough system does not have to differ too much from that in RNA world models. What is required is a population of RNA molecules, nucleotides, and amino acids; energy flow to sustain the formation and polymerization of NTP; and some form of compartmentalization to maintain sufficient concentrations of all these molecules (Fig. 2). Such compartmentalization might be realized in networks of inorganic compartments as envisaged in some recent models of early evolution [8, 13]; or in dividing lipid vesicles[81, 82]; or through a combination of these two modes. The crucial distinction of the scenario considered here from the previous ones is that the emergence of replication and translation are viewed as spontaneous events occurring in this compartmentalized primordial pool, rather than as results of (pre)biological evolution with selection. The likelihood of such complex entities emerging by chance is vanishingly small for any *O*-region but, under the "many worlds in one model", they will inevitably emerge in an infinite number of existing *O*-regions and island universes, however enormously distant from each other these "lucky" regions might be in spacetime. By virtue of anthropic selection, we live in one of such O-regions.

*The Big Bang of protein evolution*

The virtual space of protein structures (often called the protein universe[83]) is organized around ~1000 attractors named folds, i.e., distinct types of protein structure [84-87]. The proteins within the same fold, generally, are thought to share a common ancestry whereas different folds are considered to have independent evolutionary origins [86, 88]. How did the different protein folds emerge and, more generally, how did the first protein come to be, remains a mystery. Although ideas have been put forward on the possible origin of protein folds via assembly of small peptides [89], there is, currently, little empirical support for these hypotheses.

Driving, once again, from the "many worlds in one" model, which allows for a greater than previously suspected role of chance in the history of life, I propose an alternative scenario for the origin of the protein universe. I suggest that the chance convergence of events that precipitated the transition to biological evolution involved, in addition to the emergence of a primitive translation machinery and the ancestral polymerase (see above), also RNA molecules coding for the prototypes of, at least, other most widespread protein folds (the superfolds[90]). Indeed, this seems to be suggested by the "mediocrity principle" commonly applied in modern cosmology [53, 91]. Applied to the problem of the origin of proteins, this principle indicates that it is unlikely for the first polymerase mRNA to be the only RNA molecule capable of encoding a protein. Instead, the setting of the breakthrough system would include a sizable collection of potential protein-coding RNA molecules (it is conceivable that the primordial translation system did not use specific stop codons but rather relied on runaway translation – a mode that would increase the likelihood of protein coding by a random sequence). With the onset of translation, the existence of such a collection of potential protein-coding sequences would provide for explosive emergence of the prototypes of the superfolds – the "Big Bang" of protein evolution (Fig. 2). Some of the emerging proteins, e.g., RNA-binding proteins protecting RNA from hydrolysis or protoenzymes catalyzing the synthesis of RNA precursors, would facilitate the replication of the replicase mRNA and the ribosome, and accordingly, the replication of the respective mRNAs also would be selected for. This could contribute to the transition from the selection for the replication of individual RNA segments to the selection of ensembles of genetic elements (a form of group selection) which is one of the necessary earliest steps in life's evolution [13, 92, 93].

Obviously, the proposal that the major protein folds originated in a "Big Bang" event from random RNA sequences further increases the burden on chance for setting the stage of biological evolution. However, as detailed above, under the "many worlds in one model" this is not at all an insurmountable problem. Given that, this scenario provides for the rapid emergence of the RNA-protein world, in which selection-driven biological evolution would hold court, and is compatible with what we know about the structure of the protein universe.

**Discussion**

*Pinpointing the threshold of biological evolution*

The proposal made here, the spontaneous, chance origin of a RNA-protein system sufficiently complex to couple translation with replication such that biological evolution ensues, might seem extremely counterintuitive, perhaps, outrageous, even in the face of the formidable difficulty of the ORT problem. However, at least two important lines of argument could mitigate the outrage. Firstly, the postulated chance origin of the replication-translation system does not require any mysterious (let alone miraculous) processes. On the contrary, the only reactions involved are regular ones, such as polymerization of nucleotides and amino acids, nucleotide phosphorylation/dephosphorylation etc, and the only interactions required are those that are common in chemistry and biochemistry, like complementary, hydrogen-bonded interactions between polynucleotides or hydrophobic and electrostatic interactions between proteins. It is another matter that the number of such interactions postulated in this scenario is staggering, and they must be choreographed with the ultimate precision. Undeniably, the probability of such an event occurring by chance is vanishingly low. Nevertheless, the "many worlds in one" version of the eternal inflation model of the evolution of the universe has the decisive word: however unlikely any combination of physically possible events might be for any given $O$-region, in the entirety of the infinite universe, it will be encountered an infinite number of times.

The second crucial consideration pertains, more generally, to the relative contributions of chance and biological evolution to the advent of life on earth and elsewhere in the universe. Upon a more careful examination, ***any*** conceivable scenario of life's evolution necessarily requires combinations of highly unlikely conditions and events prior to the onset of biological evolution. In particular, such events include the abiogenic synthesis of fairly complex and not particularly stable organic molecules, such as nucleotides, the concentration of these molecules within appropriate compartments, and their polymerization yielding oligonucleotides of sufficient size and diversity to attain the minimal repertoire of the required catalytic activities, in particular, that of a replicase. As repeatedly emphasized here, sufficiently efficient and accurate replication is a condition *sine qua non* for biological evolution to start – everything occurring before that must result from chance. The role of chance is not eliminated by the argument that, under certain favorable conditions (e.g., in the case of nucleotides and oligonucleotides, the availability of liquid water in the appropriate range of temperatures and pH, the presence of an energy gradient that would favor phosphorylation and, perhaps, polymerization, the existence of inorganic catalysts and of natural compartments or surfaces where oligonucleotides would concentrate – the set of conditions that might exist at hydrothermal vents[8, 94, 95]), the required chemistry might become much more likely, because such conditions are rare in themselves and, again, emerge by chance.

Here I raise the possibility that the crucial transition between chance and biological evolution might lie much higher, on the scale of organizational complexity, than it is usually (explicitly or implicitly) assumed. Specifically, I propose that the most complex object that emerged by chance prior to the onset of biological evolution was a (potentially) functional, even if primitive, replication-translation system rather than a mixture of oligonucleotides, one of which would function as a low-efficiency replicase, giving rise to an RNA world (Figs. 1 and 2). In making this suggestion, I dispense of the difficult notions of the emergence of biological evolution at the stage when RNA replication, if it existed, was still inefficient (in the RNA world), gradual evolution of translation via a succession of hypothetical fitness-increasing steps (also, initially, in the RNA world), and the RNA world itself (although not of the crucial role of RNA in early evolution).

Perhaps, the strongest reason why this proposal is so counterintuitive (to the author as well) is that the postulated breakthrough system appears to be ***designed for a specific function***, such as translation: e.g., tRNAs (or their evolutionary precursors) seem to be tailored to deliver amino acids to the cognate codons in mRNAs. However, I believe that this is, largely, a semantic trap. The very notion of function makes sense only in the context of biological evolution: a complex system acquires function once the selection-driven evolution begins. Prior to that transition point, it is just a complex system, i.e., one that consists of numerous elements connected in a specific order, a system that is extremely unlikely to emerge by chance. However, as already discussed this might be irrelevant in the context of the "many worlds in one" model inasmuch as the existence of such a system does not violate any physical laws – and it obviously does not because these and even more complex systems function within life forms.

A rather subtle but critical aspect of the conceptual framework developed here is brought about by a question that is as disturbing as it is inevitable: in the redundant world of eternal inflation, why are natural selection and other mechanisms of biological evolution relevant at all? Is it not possible for any, even the highest degree of complexity to emerge purely by chance? The answer is "yes" but the question misses the point. This crucial point is that, as soon as a system capable of efficient replication emerges, natural selection sets on, and this dramatically constraints the trajectories of subsequent evolution and increases the probability of the emergence of complex life forms by orders of magnitude (Fig. 3). Thus, under the "many worlds in one" model, emergence of an infinite number of complex biotas by pure chance is inevitable but these are predicted to be vastly less common than those that evolved via the scenario that includes the switch from chance to natural selection (Fig. 3). It follows from this picture that, in any reconstruction of the origin of life, the threshold should be mapped to the lowest possible point, i.e., to the minimally complex system that is deemed capable of biological evolution. This is because, in terms of likelihood and robustness of evolutionary trajectories, biological evolution is by far superior to chance (Fig. 3), so as soon as there is an opportunity for biological evolution to start, the contribution of chance is suppressed.

In its specific, strong form, the hypothesis presented here is readily falsifiable. Indeed, as soon as the possibility of biological evolution at a low level of complexity, e.g., in the RNA world, is demonstrated and the route from the RNA world to the translation system is established, either experimentally or, at least, in a compelling model, the notion of the chance origin of a complex system with coupled processes of replication and translation will become obsolete. Falsification of this hypothesis also would be brought about by the discovery that independent origin of life is common in the universe. Although it is hard to estimate, with any reasonable accuracy, the probability of the existence of life, say, in a given galaxy, the hypothesis that the breakthrough system emerging by chance was highly complex implies extreme rarity of life. Hence it stand to reason that, e.g., the discovery of two independent instances of life within the Solar system would effectively refute the present hypothesis. However, refutation of the strong form of the present scenario via any of these routes will not affect the relevance of the "many worlds in one" model of the universe for our understanding of the origin of life. Such a discovery (tremendously important in itself) will simply lower the threshold of biological evolution on the scale of organizational complexity (Fig. 1). The general point made in this paper is that transition from chance to biological evolution is inevitable in any scenario of the origin of life. The "many worlds in one" model seems to greatly expand the span of organizational complexity that is available for this transition. One of the central goals of comparative, experimental, and theoretical research in this area is to pinpoint, as precisely as possible, the breakthrough stage (Fig. 1).

Regardless of the details, the "many worlds in one" concept changes the entire perspective on the origin and early evolution of life, and its spread in the universe. The question whether or not extraterrestrial life exists becomes obsolete: it certainly does, and in an infinite number of incarnations. In particular, there are bound to be a great variety of biospheres that function on the basis of different versions of the triplet genetic code and/or a different set of amino acids. More challenging questions could be whether or not life could evolve, i.e., biological evolution could be established on the basis of a different chemistry or the main chemical features of the life we know should be considered constraints on biological evolution. Put another way, a more general question becomes: is life, as we know it on earth, a common form or a rare aberration among the huge variety of distinct lives that are bound to exist in the infinite universe?

**Conclusions**

The emergence of replicating systems of sufficient accuracy and complexity to provide for the onset of biological evolution is a formidable problem that resists explanation within the traditional (neo)Darwinian framework. Despite considerable experimental and theoretical effort, no compelling scenarios currently exist for the origin of replication and translation, the key processes that together comprise the core of biological systems and the apparent pre-requisite for biological evolution. The RNA World concept might offer the best chance for the resolution of this conundrum but so far cannot adequately account for the emergence of an efficient RNA replicase or the translation system.

The "many worlds in one" version of the cosmological model of eternal inflation might offer a way out of this conundrum because, in an infinite universe, the spontaneous formation of complex systems by chance alone is not only possible but inevitable as long as the existence of such systems does not violate laws of physics. Thus, this model greatly expands the range of organizational complexity at which the inevitable transition from chance to biological evolution could occur in the course of life's history. Here I consider the possibility that the translation-replication system of minimal complexity required for the onset of biological evolution, along with the prototypes of the major protein folds, emerged by chance but from that point on, biological evolution became Darwinian. A crucial implication of this scenario is that, although the catalytic activity of RNA, probably, was crucial for the origin of translation, a fully fledged RNA world, with a variety of catalytic activities including that of a replicase but without a translation mechanism and encoded proteins, has never existed.

The final clarification: the ideas presented here have nothing to do with any form of "intelligent design" although, in a sense, a non-evolutionary picture of the origin of life is considered. Quite the contrary, and regardless of where, precisely, the threshold of biological evolution is placed on the timeline of life's history, I believe that placing the role of chance in the origin of life in the context of the cosmic necessity serves to demystify this enigmatic but momentous event.

**Acknowledgments**

I thank Alex Vilenkin for critical reading of the manuscript and constructive suggestions, and Tania Senkevich for stimulating discussions.

Figure legends

**Figure 1**. The transition from chance to biological evolution in the history of life.

The grey area and dotted lines illustrate the uncertainty in the identification of the transition point, i.e., the level of complexity at which the transition occurred.

**Figure 2**. The emergence of the breakthrough system from within a primordial pool of RNA molecules.

For the sake of concreteness, the figure employs the specific model of the evolution of pre-cellular life within networks of inorganic compartments that was discussed in detail in a series of previous publications [8, 13, 17, 96]. The arrows connecting the compartments denote the extensive transfer of content between the compartments that is salient to the model . However, the scenario considered here is conceived as a general one and is not conditional on this particular model of early evolution.

**Figure 3**. The narrowing of the range of possible histories and the increased likelihood of the emergence of high complexity brought about by the transition form chance to biological evolution.

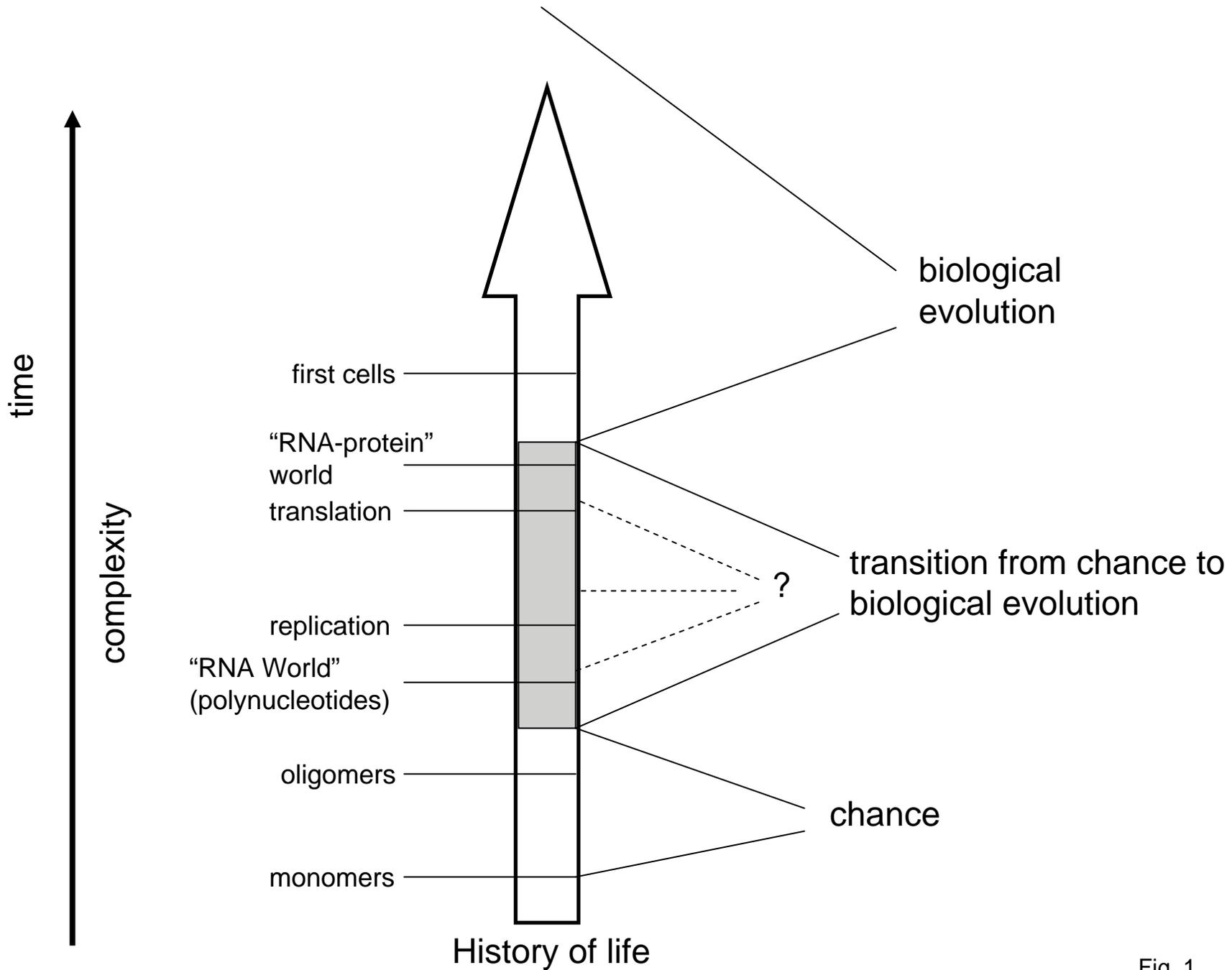

Fig. 1

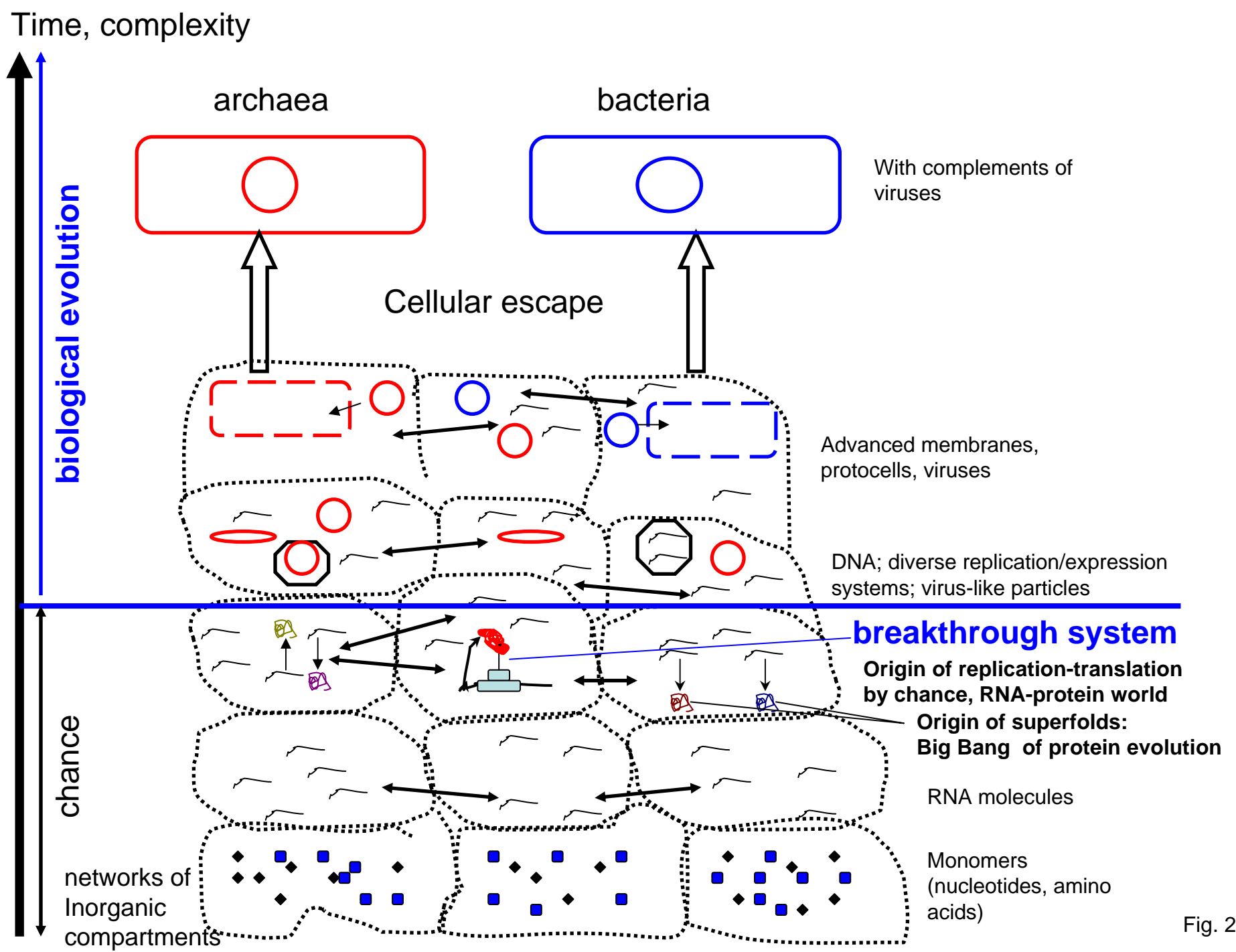

Fig. 2

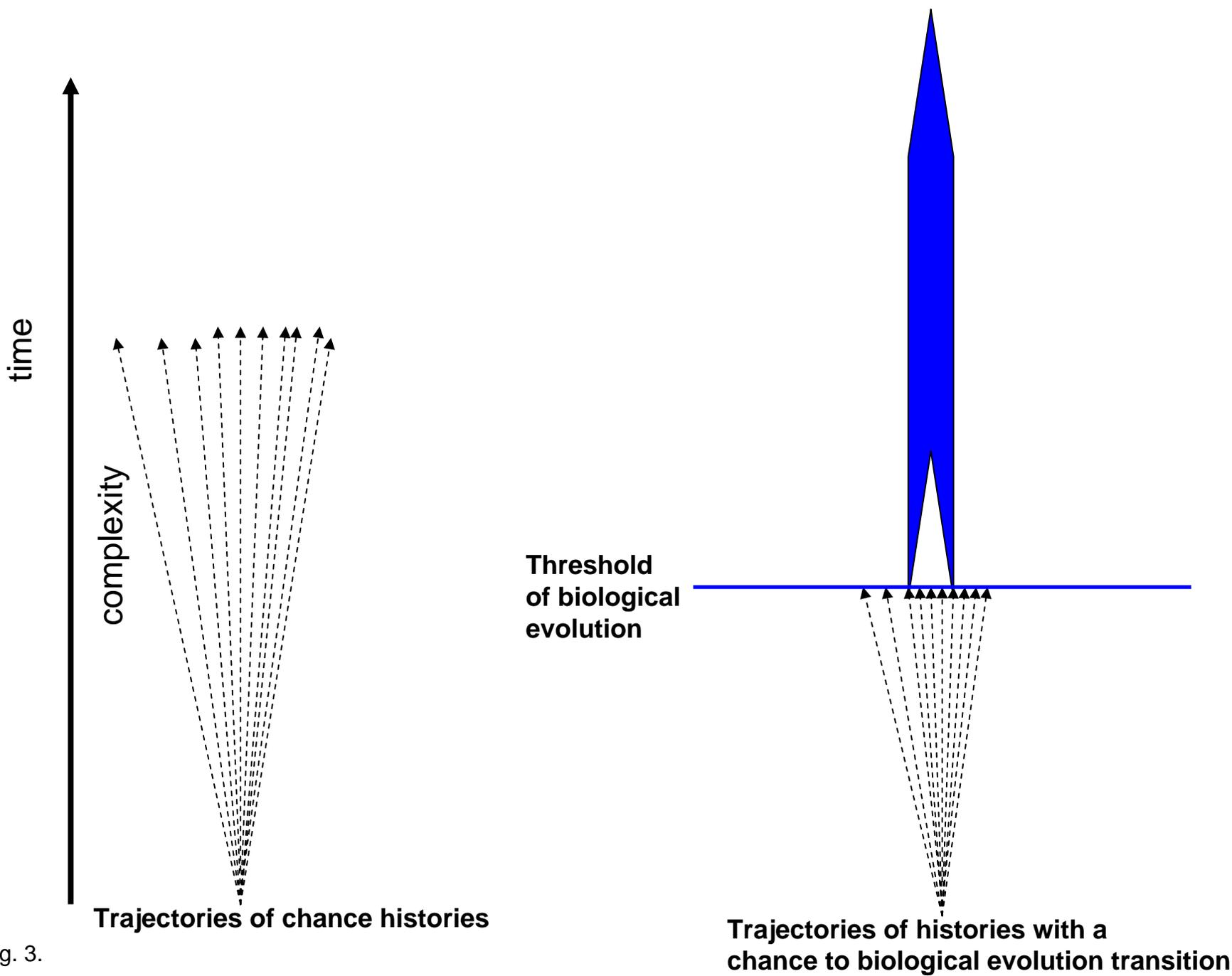

Fig. 3.